\PassOptionsToPackage{dvipsnames,svgnames}{xcolor}

\documentclass[sigconf,screen]{aamas} 

\usepackage{balance} 

\usepackage[nameinlink,capitalise]{cleveref} 

\usepackage{makecell}
\usepackage{multirow}

\newcommand{\our}{{\textsc{CAMP}}}

\usepackage{fontawesome}

\makeatletter
\gdef\@copyrightpermission{
  \begin{minipage}{0.2\columnwidth}
   \href{https://creativecommons.org/licenses/by/4.0/}{\includegraphics[width=0.90\textwidth]{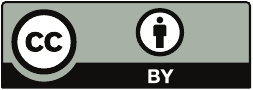}}
  \end{minipage}\hfill
  \begin{minipage}{0.8\columnwidth}
  This work is licensed under a Creative Commons Attribution International 4.0 License.
  \end{minipage}
  \vspace{5pt}
}
\makeatother

\setcopyright{ifaamas}
\acmConference[AAMAS '25]{Proc.\@ of the 24th International Conference
on Autonomous Agents and Multiagent Systems (AAMAS 2025)}{May 19 -- 23, 2025}
{Detroit, Michigan, USA}{Y.~Vorobeychik, S.~Das, A.~Nowé  (eds.)}
\copyrightyear{2025}
\acmYear{2025}
\acmDOI{}
\acmPrice{}
\acmISBN{}

\acmSubmissionID{1389}

\def \papertitle{CAMP: Collaborative Attention Model with Profiles for Vehicle Routing Problems}

\title[\papertitle]{\papertitle}

\author{Chuanbo Hua$^*$}
\affiliation{
  \institution{KAIST, Omelet}
  \city{Daejeon}
  \country{South Korea}}
\email{cbhua@kaist.ac.kr}

\author{Federico Berto$^*$}
\affiliation{
  \institution{KAIST, Omelet}
  \city{Daejeon}
  \country{South Korea}}
\email{fberto@kaist.ac.kr}

\author{Jiwoo Son$^*$}
\affiliation{
  \institution{Omelet}
  \city{Busan}
  \country{South Korea}}
\email{jiwoo.son@omelet.ai}

\author{Seunghyun Kang}
\affiliation{
  \institution{Omelet}
  \city{Daejeon}
  \country{South Korea}}
\email{seunghyun.kang@omelet.ai}

\author{Changhyun Kwon}
\affiliation{
  \institution{KAIST, Omelet}
  \city{Daejeon}
  \country{South Korea}}
\email{chkwon@kaist.ac.kr}

\author{Jinkyoo Park}
\affiliation{
  \institution{KAIST, Omelet}
  \city{Daejeon}
  \country{South Korea}}
\email{jinkyoo.park@kaist.ac.kr}

\thanks{$^*$These authors contributed equally to this work. Work made with contributions from the AI4CO open research community.}

\begin{abstract}
The profiled vehicle routing problem (PVRP) is a generalization of the heterogeneous capacitated vehicle routing problem (HCVRP) in which the objective is to optimize the routes of vehicles to serve client demands subject to different vehicle profiles, with each having a preference or constraint on a per-client basis. 
While existing learning methods have shown promise for solving the HCVRP in real-time, no learning method exists to solve the more practical and challenging PVRP. 
In this paper, we propose a Collaborative Attention Model with Profiles (\our{}), a novel approach that learns efficient solvers for PVRP using multi-agent reinforcement learning. \our{} employs a specialized attention-based encoder architecture to embed profiled client embeddings in parallel for each vehicle profile. We design a communication layer between agents for collaborative decision-making across profiled embeddings at each decoding step and a batched pointer mechanism to attend to the profiled embeddings to evaluate the likelihood of the next actions. We evaluate \our{} on two variants of PVRPs: PVRP with preferences, which explicitly influence the reward function, and PVRP with zone constraints with different numbers of agents and clients, demonstrating that our learned solvers achieve competitive results compared to both classical state-of-the-art neural multi-agent models in terms of solution quality and computational efficiency. We make our code openly available at \url{https://github.com/ai4co/camp}.

\end{abstract}


\newcommand{\BibTeX}{\rm B\kern-.05em{\sc i\kern-.025em b}\kern-.08em\TeX}

\usepackage{amsmath, amsfonts, mathtools,bm}
\usepackage{physics}
\usepackage{cancel}
\usepackage{thmtools, thm-restate}
\usepackage{accents}

\usepackage{pict2e, picture}
\makeatletter
\DeclareRobustCommand{\Arrow}[1][]{%
\check@mathfonts
\if\relax\detokenize{#1}\relax
\settowidth{\dimen@}{$\m@th\rightarrow$}%
\else
\setlength{\dimen@}{#1}%
\fi
\sbox\z@{\usefont{U}{lasy}{m}{n}\symbol{41}}%
\begin{picture}(\dimen@,\ht\z@)
\roundcap
\put(\dimexpr\dimen@-.7\wd\z@,0){\usebox\z@}
\put(0,\fontdimen22\textfont2){\line(1,0){\dimen@}}
\end{picture}%
}
\makeatother

\DeclareMathAlphabet{\nummathbb}{U}{BOONDOX-ds}{m}{n}

\makeatletter
\DeclareRobustCommand\widecheck[1]{{\mathpalette\@widecheck{#1}}}
\def\@widecheck#1#2{%
    \setbox\z@\hbox{\m@th$#1#2$}%
    \setbox\tw@\hbox{\m@th$#1%
       \widehat{%
          \vrule\@width\z@\@height\ht\z@
          \vrule\@height\z@\@width\wd\z@}$}%
    \dp\tw@-\ht\z@
    \@tempdima\ht\z@ \advance\@tempdima2\ht\tw@ \divide\@tempdima\thr@@
    \setbox\tw@\hbox{%
       \raise\@tempdima\hbox{\scalebox{1}[-1]{\lower\@tempdima\box
\tw@}}}%
    {\ooalign{\box\tw@ \cr \box\z@}}}
\makeatother

\begin{document}



\hypersetup{
  bookmarks=true,
  colorlinks=true,
  linkcolor=magenta,
  citecolor=ForestGreen,
  urlcolor=blue!80!black,
  filecolor=blue!80!black
  }


\pagestyle{fancy}
\fancyhead{}


\maketitle 


\section{Introduction}

Vehicle Routing Problems (VRPs) are a class of well-known combinatorial optimization (CO) problems where the objective is to determine the most efficient set of routes for a fleet of vehicles to deliver goods to a set of clients. A particularly challenging variant is the Heterogeneous Capacitated Vehicle Routing Problem (HCVRP), where the fleet consists of vehicles with varying capacities and operational costs \citep{golden1984fleet}. In many real-world scenarios, different vehicles might not only have capacity constraints but also profile-specific preferences or operational constraints that can vary per client. This problem, which we term the Profiled Vehicle Routing Problem (PVRP), generalizes HCVRP by incorporating vehicle profiles with client-specific preferences or constraints \citep{braekers2016vehicle}. These profiles might affect routing decisions based on factors like vehicle access to particular areas, preferred client relationships, or regulatory requirements for specific vehicle-client combinations \citep{zhong2007territory, aiko2018incorporating, locus2020zonebased, li2023experience}. Solving the PVRP with exact solution methods, such as Branch and Bound, becomes impractical for large instances due to its NP-hard nature \citep{papadimitriou1998combinatorial}. Thus, heuristic approaches like genetic algorithms are used in practical scenarios offering approximate solutions trying to balance solution quality and computational efficiency, but they may struggle in larger scales \citep{lozano2011editorial} and require considerable manual design and tuning \citep{johnson1997traveling}.

Recent advances in reinforcement learning (RL) for combinatorial optimization provide a promising alternative with several benefits. RL can automatically learn effective solutions without supervision by interacting with relatively inexpensive simulated environments; thus, designing RL methods requires little to no domain expertise. RL methods can find faster and possibly better solutions than heuristics approaches, particularly in improving scalability to more complex real-world problem instances. Most works follow the pointer network paradigm \citep{vinyals2015pointer, bello2016neural}, an encoder-decoder architecture that encodes input data into a shared latent space, which is then used for fast autoregressive decoding \citep{sun2019fast}.


Although recent learning-based approaches have been applied successfully to various VRPs \citep{kool2018attention, kwon2020pomo, kim2022sym,luo2023neural}, including rich and highly constrained variants \citep{zhou2023_lns_vrp_aph, liu2024multi, zhou2024mvmoe, jiang2024unco, berto2024routefinder, bi2024learning, li2024cada}, multi-agent \citep{ zhang2020multi, falkner2020learning, zong2022mapdp, park2023learn, son2024equity,zheng2024dpn,zheng2024udc,gama2024multi_agent_envs_vrp, deineko2024learn_asap_time_constrained} and heterogenous agent (i.e., fleet) VRPs \citep{li2021heterogeneous, liu20242d,berto2024parco}, such approaches cannot model heterogeneous agents on a \textit{per-client} basis, in which each agent has a different internal representation for each node.

In this paper, we propose a novel learning approach, the Collaborative Attention Model with Profiles (\our{}), designed to address PVRP using multiagent reinforcement learning (MARL). Our approach builds upon the attention-based encoder-decoder framework \citep{kool2018attention}, integrating vehicle and client profiles into a collaborative decision-making architecture. Unlike previous methods, \our{} leverages a specialized attention-based communication encoder to embed profiled client representations in parallel for each vehicle. During the decoding phase, we employ a specialized communication layer between agents to enable cooperative decisions even among a heterogeneous profiled latent space. Finally, a parallel pointer mechanism \citep{berto2024parco} is utilized to attend to these profiled embeddings, allowing the model to efficiently evaluate possible next actions for each vehicle based on its profile.

We evaluate \our{} on the PVRP with Preferences (PVRP-P), where various distributions of vehicle profiles explicitly affect the reward function, and PVRP with Zone Constraints (PVRP-ZC), a scenario that imposes zone-based operational constraints on vehicles.
Our experiments demonstrate that \our{} achieves competitive results in terms of solution quality and computational efficiency compared to classical methods and modern neural multi-agent models, showcasing \our{} as a valuable tool for research and practitioners in solving complex routing problems in real-time.


\section{Related Work}

Recent advancements in neural combinatorial optimization (NCO) have introduced promising end-to-end solutions for vehicle routing problems (VRP) \citep{bengio2021_ml4co_survey,yang2023_rl4co_survey}. Learning-based VRP approaches can broadly be divided into \textit{construction} \citep{kool2018attention,kwon2020pomo, kim2022sym,bogyrbayeva2023deep,
grinsztajn2023winner,hou2023generalize, jin2023pointerformer,gao2023_transferrable_local_policy_vrp, xiao2023_nar_solver_tsp, luo2023neural, jiang2023ensemble_vrp, li2024distribution, drakulic2024bq, drakulic2024goal, lin2024cross, zhou2024collaboration} and \textit{improvement/search} methods \citep{hottung2021efficient,ma2021learning_dact,son2023meta_sage,sun2023difusco, ma2024learning, ye2024glop, ye2023deepaco, kim2024gfacs, verdu2024scaling_meta_sage_2,hottung2025neural}. Construction methods learn to generate a solution, while improvement/search methods iteratively refine them. We focus on learning construction methods due to their lower reliance on handcrafted heuristics and faster inference speed. Most constructive methods derive from the seminal work of \citet{vinyals2015pointer}, refined by \citet{bello2016neural} with deep reinforcement learning, and improved by \citet{kool2018attention} using transformers in the widely adopted Attention Model (AM). More practical multi-agent VRPs have been formulated by \citet{son2024equity}and \citet{zheng2024dpn} as sequential decision-making, while \citet{zhang2020multi} and \citet{zong2022mapdp} and \citet{liu20242d} employ simultaneous solution construction. For the heterogeneous VRP (HCVRP), approaches have evolved from \citet{vera2019deep}'s policy network with A2C method to \citet{qin2021novel}'s reinforcement learning-enhanced heuristics and \citet{li2022deep}'s two-decoder framework emphasizing vehicle and node selection. However, these methods often lack fleet generalization due to fixed vehicle number assumptions, and while they attempt simultaneous solutions for various agents, they typically sequentially limit fast parallel decoding and collaboration. PARCO \citep{berto2024parco} addresses these limitations with a flexible approach using a general communication framework, making it effective for multi-agent and heterogeneous VRP scenarios, enhancing solutions by allowing for dynamic agent counts and improved conflict resolution strategies, representing a significant advancement in addressing the complexities of HCVRPs within the NCO framework. However, no neural approach has yet been proposed to tackle the more practical PVRP.

\section{Problem Formulation}
\label{sec:preliminaries}

\subsection{Mathematical Formulation of PVRP}

Consider a set of nodes $N = \{0, 1, \ldots, n\}$, where node 0 represents the depot and nodes $\{1, \ldots, n\}$ represent clients. Let $C = N \setminus \{0\}$ denote the set of client nodes. The set of vehicles is represented by $K = \{1, \ldots, m\}$. Each node $i \in N$ is characterized by its location $s_i \in \mathbb{R}^2$ and demand $d_i$ (with $d_0 = 0$ for the depot). Each vehicle $k \in K$ is defined by its capacity $Q_k$, speed $s_k$, and a profile parameter $p_k = (p_{1k}, \ldots, p_{nk})$, where $p_{ik}$ represents the profile parameter for vehicle $k$ serving client $i$. Let $r$ denote index for delivery trips, $r\in R$, where $R$ is the maximum number of trips for each vehicle.

Let $c_{ij}$ denote the travel cost from node $i$ to node $j$, typically representing the Euclidean distance between the nodes. The decision variable is a boolean
$x^r_{ijk}$ that equals $1$ if vehicle $k$ travels directly from node $i$ to node $j$ in trip $r$, and 0 otherwise. We also introduce three auxiliary variables to aid in problem formulation and solution: $y^r_{ik}$ is a binary variable that equals $1$ if client $i$ is served by vehicle $k$ in trip $r$, and 0 otherwise. $z_k$ is a binary variable representing the vehicle $k$ is selected or not. Let $w_i$ be a continuous variable used to avoid subtours.

The PVRP can thus be formulated as follows:

\begin{equation}
\min \sum_{k \in K} \sum_{i \in N} \sum_{j \in N} \sum_{r \in R} \left(\frac{c_{ij}}{s_k} - \alpha p_{ik}\right) x^r_{ijk}
\label{eq:objective}
\end{equation}

\noindent
subject to the following constraints:

\begin{align}
& \sum_{k \in K} \sum_{r \in R} y^r_{ik} = 1, && \forall i \in C \label{eq:visit_once} \\
& \sum_{j \in C} x^r_{0jk} \leq z_k, && \forall k \in K, r \in R \label{eq:leave_depot} \\
& \sum_{j \in N} x^r_{ijk} \leq y^r_{ik}, && \forall i \in N, k \in K, r \in R \label{eq:linking_constraint} \\
& x^r_{iik} = 0, && \forall k \in K, r \in R \label{eq:ban_same_location} \\
& \sum_{i \in N} x^r_{ihk} - \sum_{j \in N} x^r_{hjk} = 0, && \forall h \in N, k \in K, r \in R \label{eq:flow_conservation} \\
& \sum_{j \in C} x^r_{0jk} = \sum_{i \in C} x^r_{i0k}, && \forall k \in K, r \in R \label{eq:depot_start_end} \\
& \sum_{i \in C} \sum_{j \in C} d_j x^r_{ijk} \leq Q_k, && \forall i,j \in N, k \in K, r \in R \label{eq:capacity_respect} \\
& w_j \geq w_i + 1 - N(1 - \sum_k x^r_{ijk}), && \forall i \in C, j \in C, i \neq j, r \in R \label{eq:subtour_elimination} \\
& \sum_{j \in C} x^r_{0jk} >= \sum_{j \in C} x^{r+1}_{0jk}, && \forall k \in K, r \leq R-1 \label{eq:trip_sequence} \\
& x^r_{ijk}, y^r_{ik}, z_k \in \{0,1\}, && \forall i,j \in N, k \in K, r \in R \label{eq:binary_constraint}
\end{align}
The objective function \cref{eq:objective} minimizes the total adjusted travel cost while maximizing preference scores, with $\alpha$ as a weight parameter to balance these two factors. Constraint \eqref{eq:visit_once} ensures that each client is visited exactly once. Constraint \eqref{eq:leave_depot}
counts utilized vehicle while leaving from depot. Constraint \eqref{eq:linking_constraint} links the route variables ($x^r_{ijk}$) with the assignment variables ($y^r_{ik}$). Constraint \eqref{eq:ban_same_location} bans traveling in the same location. Constraint \eqref{eq:flow_conservation} maintains flow conservation for each vehicle, while constraint \eqref{eq:depot_start_end} guarantees that each vehicle forces that if a vehicle leaves the
depot, it must return to the depot. Constraint \eqref{eq:capacity_respect} ensures that vehicle capacities are respected. Constraint \eqref{eq:subtour_elimination} eliminates subtours. Constraints \eqref{eq:trip_sequence} establishes
trip sequences. Constraints \eqref{eq:binary_constraint} define the binary for the decision variables.


\subsubsection{PVRP with Preferences}
\label{subsec:prvp-p-formulation}

The PVRP-P defines the profile parameter $p_{ik}$ as a \textit{preference score}. The objective function \cref{eq:objective} can be simply rewritten as follows:

\begin{equation}
\max \sum_{k \in K} \sum_{i \in N} \sum_{j \in N} \left(\alpha p_{ik} - \frac{c_{ij}}{s_k} \right) x_{ijk},
\label{eq:objective_vrp-p}
\end{equation}

\noindent
i.e., maximize the preferences while minimizing the total tour duration with a multi-objective balancing parameter $\alpha$.

\subsubsection{PVRP with Zone Constraints}
\label{subsec:prvp-zc-formulation}

The PVRP-ZC incorporates zone constraints, where certain vehicles may not serve specific clients. One way to model this problem is to introduce an additional constraint:
\begin{equation}
y_{ik} = 0, \quad \forall (i,k) \text{ where } p_{ik} = 0.
\end{equation}
In practice, this constraint can be modeled by \cref{eq:objective} by introducing a large negative value, effectively $-\infty$\footnote{In practice, a sufficiently large (negative) constant $M$ can be chosen to avoid numerical overflows. For instance, OR-Tools sets $|M|=2^{48}$.}, for $p_{ik}$ when vehicle $k$ is not allowed to serve client $i$:
\begin{equation}
p_{ik} = \begin{cases}
-\infty & \text{if vehicle } k \text{ is not allowed to serve client } i \\
0 & \text{otherwise}
\end{cases}
\end{equation}
and setting $\alpha$ to $1$. Using this definition of $p_{ik}$, the objective function \cref{eq:objective_vrp-p} naturally enforces zone constraints. When $p_{ik} = -\infty$, the term $- \alpha p_{ik}$ becomes $\infty$, ensuring that the corresponding $x_{ijk}$ variables are set to 0 in any optimal solution, effectively preventing vehicle $k$ from serving client $i$.

\begin{figure}[h!]
    \centering
    \includegraphics[width=\linewidth]{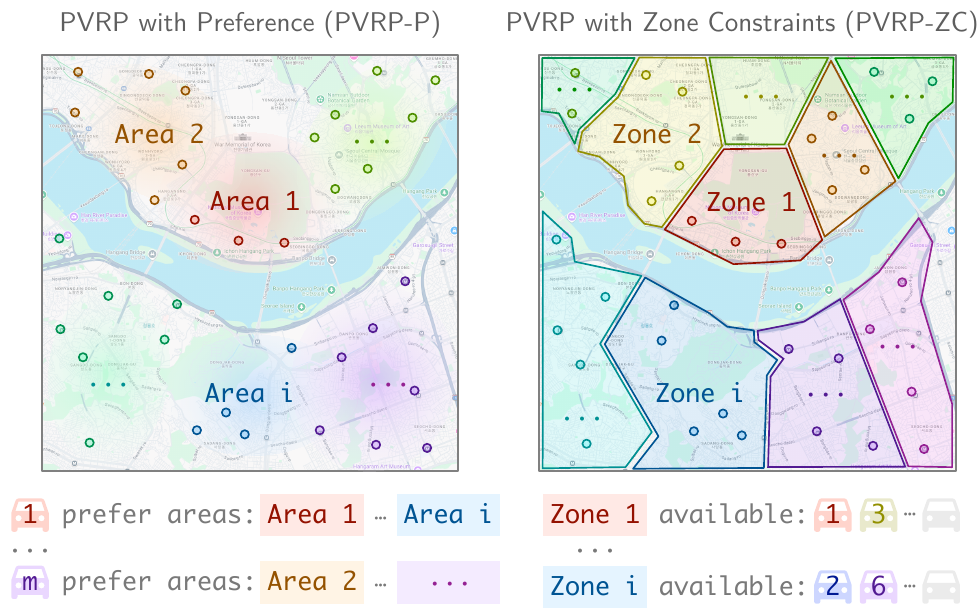}
    \vspace{-5mm}
    \caption{Practical examples of PVRPs. [Left] PVRP-P has preference zones for each vehicle profile. [Right] PVRP-ZC has (hard) zone constraints for certain vehicles.}
    \label{fig:pvrp-catchy}
\end{figure}

We show practical examples of PVRP-P (\cref{subsec:prvp-p-formulation}) and PVRP-ZC (\cref{subsec:prvp-zc-formulation}) in \cref{fig:pvrp-catchy}.


\section{Methodology}
\label{sec:methodology}

\subsection{Modeling PVRP with MARL}
\label{subsec:pvrp-marl-mdp-formulation}

We reformulate the Profiled Vehicle Routing Problem (PVRP) as a Markov Decision Process (MDP) defined by the 4-tuple $M = \{S, A, \tau, r\}$, where $S$ is the state space, $A$ is the action space, $\tau$ is the state transition function, and $r$ is the reward function. 

\paragraph{\textbf{State Space $S$}}
Each state $s_t = (V_t, X_t) \in S$ consists of two parts:

1. Vehicle state $V_t = \{v_t^1, v_t^2, \ldots, v_t^m\}$, where each $v_t^k = (o_t^k, T_t^k, G_t^k, p_k)$ represents:
   \begin{itemize}
   \item $o_t^k$: Remaining capacity of vehicle $k$ at step $t$
   \item $T_t^k$: Accumulated travel time of vehicle $k$ at step $t$
   \item $G_t^k = \{g_0^k, g_1^k, \ldots, g_t^k\}$: Partial route of vehicle $k$ at step $t$
   \item $p_k = (p_{1k}, \ldots, p_{nk})$: Profile parameter of vehicle $k$
   \end{itemize}

   The interpretation of $p_k$ differs between PVRP-P and PVRP-ZC:
   \begin{itemize}
   \item For PVRP-P: $p_{ik}$ represents the preference score for vehicle $k$ serving node $i$
   \item For PVRP-ZC: $p_{ik}$ is binary, where $1$ indicates vehicle $k$ is allowed to serve node $i$, and $0$ indicates it is not allowed
   \end{itemize}

2. Node state $X_t = \{x_t^0, x_t^1, \ldots, x_t^n\}$, where each $x_t^i = (s^i, d_t^i)$ represents:
   \begin{itemize}
   \item $s^i$: 2D vector representing the location of node $i$
   \item $d_t^i$: Remaining demand of node $i$ at step $t$
   \end{itemize}

\paragraph{\textbf{Action Space $A$}}
Without loss of generality, we consider an action $a_t \in A$ is defined as selecting a vehicle and a node to visit. Specifically, $a_t = (v_t^k, x_t^j)$, where vehicle $k$ is selected to visit node $j$ at step $t$.

\paragraph{\textbf{State Transition Function $\tau$}}
The transition function $\tau$ updates the state based on the chosen action:
$s_{t+1} = (V_{t+1}, X_{t+1}) = \tau(V_t, X_t, a_t)$.
For PVRP-P, the elements are updated as follows:

\begin{equation}
o_{t+1}^k = \begin{cases}
o_t^k - d_t^j, & \text{if } k \text{ is the selected vehicle} \\
o_t^k, & \text{otherwise}
\end{cases}
\end{equation}

\begin{equation}
T_{t+1}^k = \begin{cases}
T_t^k + \frac{c_{g_t^k,j}}{s_k}, & \text{if } k \text{ is the selected vehicle} \\
T_t^k, & \text{otherwise}
\end{cases}
\end{equation}

\begin{equation}
G_{t+1}^k = \begin{cases}
[G_t^k, j], & \text{if } k \text{ is the selected vehicle} \\
G_t^k, & \text{otherwise}
\end{cases}
\end{equation}

\begin{equation}
d_{t+1}^i = \begin{cases}
0, & \text{if } i \text{ is the selected node} \\
d_t^i, & \text{otherwise}
\end{cases}
\end{equation}

For PVRP-ZC, we modify the transition function to incorporate zone constraints:

\begin{equation}
\tau_{ZC}(V_t, X_t, a_t) = \begin{cases}
(V_{t+1}, X_{t+1}), & \text{if } p_{jk} = 1 \\
(V_t, X_t), & \text{if } p_{jk} = 0
\end{cases}
\end{equation}
i.e., actions for all vehicle $k$ which would lead $p_{jk} = 0$ are directly masked in the environment.

Where $(V_{t+1}, X_{t+1})$ is calculated using the same update rules as in PVRP-P. This ensures that the state remains unchanged if an invalid action is attempted.

\paragraph{\textbf{Reward Function $r$}}
The reward function is defined as follows for both PVRP-P and PVRP-ZC:

\begin{equation}
r_t = \begin{cases}
0, & \text{if episode is not complete} \\
R(s_T), & \text{if episode is complete (at final step T)}
\end{cases}
\end{equation}

Where $R(s_T)$ is the final reward calculated at the end of the episode:

for PVRP-P:
\begin{equation}
R(s_T) = \sum_{k \in K} \sum_{(i,j) \in G_T^k} \left(\alpha p_{ik} - \frac{c_{ij}}{s_k}\right)
\end{equation}

for PVRP-ZC:
\begin{equation}
R(s_T) = -\sum_{k \in K} \sum_{(i,j) \in G_T^k} \frac{c_{ij}}{s_k}
\end{equation}

Where $G_T^k$ is the final route of vehicle $k$, $c_{ij}$ is the travel cost between nodes $i$ and $j$, $s_k$ is the speed of vehicle $k$, $p_{ik}$ is the preference score for vehicle $k$ serving node $i$, and $\alpha$ is a weight parameter balancing travel cost and preference score (for PVRP-P only).

\subsubsection{Optimization Objective}

The goal is to find the optimal policy parameters $\theta^*$ that maximize the expected cumulative reward. Formally, we aim to solve:

\begin{equation}
\theta^* = \arg\max_\theta \mathbb{E}_{\tau \sim \pi_\theta} \left[\sum_{t=0}^T r_t\right]
\end{equation}

\subsubsection{Construction Methods}
\label{subsec:construction-methods}

We note that there are several ways to construct solutions, starting from the formulated MDP, which takes into account the most general case. In particular, we identify the following cases:

\begin{enumerate}
    \item \textit{Autoregressive Sequential}: such construction method considers a single vehicle and one action at a time; a vehicle is switched only when its route is complete \citep{son2024equity,zheng2024dpn}.
    \item \textit{Autoregressive Alternating}: similar to the above but more flexible: in this case, the vehicle can be switched at any time \citep{li2021heterogeneous,liu20242d}.
    \item \textit{Parallel Autoregressive}:  in this case, the steps are performed for all agents in parallel simultaneously, reducing the total $T$, which makes optimization and inference faster \citep{berto2024parco}.
\end{enumerate}

Without loss of generality, we adopt the Parallel Autoregressive approach in \our{} because of its speed and flexibility.

\begin{figure*}[ht!]
    \centering
    \includegraphics[width=0.88\linewidth]{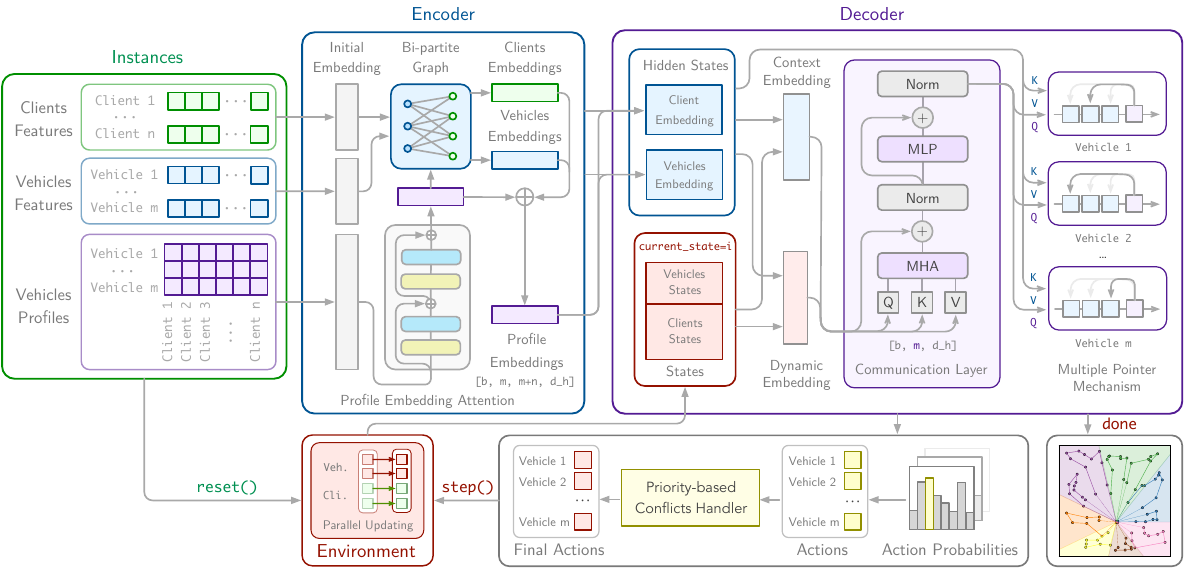}
    \caption{Overview of \our{}.}
    \label{fig:camp-model-catchy}
\end{figure*}

\subsection{Collaborative Attention Model with Profiles}
\label{subsec:main-camp}

\cref{fig:camp-model-catchy} shows an illustration of \our{} model. We first provide an overview of the overall solution construction (\cref{subsec:solution-construction-camp}), then of the encoder (\cref{subsec:encoder-camp}) and decoder (\cref{subsec:decoder-camp}).

\subsubsection{Overall solution construction}
\label{subsec:solution-construction-camp}

We formulate the solution construction of \our{} in a parallel autoregressive approach \citep{berto2024parco}.
%
%
We define the solution construction $\bm{a}$ given instance $\bm{x}$ as follows:
\begin{align}
\label{eq:parco_encoding_decoding}
\bm{h} &= f_{\theta}(x) \\
\pi_\theta(\bm{a}|\bm{x}) &= \prod_{t=0}^T  \prod_{k=1}^{m} g_\theta(a_{t}^k | a_{t-1}^k, \dots, a_{0}^k, \bm{h})
\end{align}
where $f_\theta(\cdot)$ is the \textit{encoder} $f$ mapping $\bm{x}$ to its (vehicle-specific) latent embeddings $\bm{h}$, $g_{\theta}(\cdot)$ is the autoregressive \textit{decoder}, and $\pi_\theta$ represents the full \our{} encoder-decoder model mapping $\bm{x}$ to $\bm{a}$.

\subsubsection{Encoder}
\label{subsec:encoder-camp}

\paragraph{Encoder}
The encoder in \our{} is designed to handle multiple vehicle profiles, each with distinct embeddings. Unlike previous approaches that utilize a single shared embedding for all vehicles \citep{son2024equity,zheng2024dpn,berto2024parco}, \our{} encodes each vehicle profile individually, generating profile-specific embeddings. The encoding process involves three key steps: (1) obtaining initial embeddings, (2) applying attention to compute vehicle-specific profile embeddings, and (3) integrating these profile embeddings using bipartite graph message passing. 

First, the encoder maps the raw features of the graph instance into an embedding space. Specifically, it transforms the feature vectors of vehicles \( x^{v} \), clients \( x^{c} \), and vehicle preference profiles \( p_{ij} \) into their respective embedding spaces using the learned embedding matrices \( W_{\text{init}}^{v} \), \( W_{\text{init}}^{c} \), and \( W_{\text{init}}^{p} \). The resulting embeddings are computed as follows:
\begin{align*}
    h^{v}_i &= W_{\text{init}}^{v} \cdot x^{v}_i \\
    h^{c}_j &= W_{\text{init}}^{c} \cdot x^{c}_j \\
    h^{p}_{ij} &= W_{\text{init}}^{p} \cdot p_{ij}
\end{align*}

Next, the vehicle, client, and preference embeddings are concatenated to form a combined profile embedding, integrating both node and edge features in the graph:
\begin{align*}
    h_{ij} &= W_{\text{combine}} \cdot \text{concat}(h_i^{v}, h_j^{c}, h_{ij}^{p})
\end{align*}

This profile embedding \( h_{ij} \), which captures the interaction between a vehicle and clients, is processed using multi-head attention (MHA) to incorporate profile-specific information. The overall profile embedding \( \bm{h} \) is constructed by concatenating the embeddings of all vehicle-client pairs: \( \bm{h} = \text{concat}(h_{1}, h_{2}, \dots, h_{m}) \). MHA is then applied separately to each profile:
\begin{align}
    h^\prime &= \text{MHA}(h_{i}, h_{i}, h_{i})
\end{align}
where MHA represents the multi-head attention defined as:
\begin{align*}
    \text{MHA}(Q, K, V) &= \left( \mathop{\big\|}_{i=1}^{n_\text{heads}} \text{Attention}(QW^Q_i, KW^K_i, VW^V_i) \right) W^O 
\end{align*}
with
\begin{align*}
    \text{Attention}(Q, K, V) &= \text{softmax} \left( \frac{QK^\top}{\sqrt{d_k}} \right) V \,
\end{align*}
where \( \mathop{\big\|} \) denotes concatenation; \( W^O \in \mathbb{R}^{d_k \times d_h} \) is used to combine the outputs from different attention heads where \( d_k = d_h / {n_\text{heads}} \) represents the dimension per heads; and \( W^Q_i, W^K_i, W^V_i \in \mathbb{R}^{d_h \times d_k} \) are the learnable parameter matrices for queries \( Q \), keys \( K \), and values \( V \). 

After obtaining the vehicle-specific profile embeddings, the encoder in \our{} employs a bipartite graph message passing framework, which is crucial for integrating information across individual vehicle profile embeddings. By facilitating the exchange of information between vehicle and client nodes, this framework allows each vehicle-specific profile embedding to benefit from the broader context, enabling a richer, more interconnected representation. The bipartite graph structure enables two-directional message passing:

\begin{itemize}
    \item \textbf{Vehicle-to-Client Update}: The vehicle embeddings are propagated to the client nodes they are connected to, aggregating vehicle-specific information to update customer embeddings:
    \[
    h'^{c}_{j} = \Phi(h^{v}_i, h^{c}_j, h'_{ij})
    \]

    \item \textbf{Client-to-Vehicle Update}: The customer embeddings are subsequently propagated back to the vehicles, allowing vehicles to gather information from multiple customers:
    \[
    h'^{v}_{i} = \Phi(h^{c}_{j}, h^{v}_i, h'_{ji})
    \]
\end{itemize}

This bidirectional message passing ($\Phi$) enables information to flow across all vehicle-customer pairs, allowing vehicle-specific profile embeddings to interact and integrate with each other.

Residual connections are employed to maintain stability in the learning process, which add the initial node embeddings to the updated embeddings:
\[
h''^{v}_i = h^{v}_i + h'^{v}_{i}, \quad h''^{c}_j = h^{c}_j + h'^{c}_{j}
\]
Finally, the edge embeddings are refined by combining the updated vehicle and customer embeddings, ensuring that the edge representations reflect the fully integrated node information:
\[
h''_{ij} = \text{concat}(h^{v}_i, h^{c}_j) + h'_{ij}
\]

The aggregated information is then passed through multiple transformer-style layers \citep{vaswani2017attention} with bidirectional message passing similarly to \cref{eq:transformer-layer}. The final output of the encoder is a set of embeddings $\bm{h} = [\bm{h}^{1}, \ldots, \bm{h}^{m}]$, where each $\bm{h}^k \in \mathbb{R}^{(m+n) \times d_h}$ represents the encoded graph of vehicles and clients for vehicle profile $k$. This approach allows \our{} to capture profile-specific information and relationships between vehicle and client, which is crucial for solving the PVRP. 



\subsubsection{Decoder}
\label{subsec:decoder-camp}

The decoder transforms multi-profile embeddings $\bm{h}$ into a probability distribution for action selection. 
We calculate the $m$ queries $Q_t$ for the multiple pointer mechanism \citep{berto2024parco} as follows:
\begin{equation}
    Q_t = W_\text{proj}(\text{concat}(\bm{h}^k_k, \bm{h}^k_i, W_\text{context} \cdot x_t))
\end{equation}
where $k$ is the vehicle index, $i$ is the node index, $x_t$ are context features at the current step $t$; $W_\text{context}$ is a learnable parameter matrix to project the context features while $W_\text{proj}$ is a learnable parameter matrix to project back to hidden space $d_h$ the concatenated static embeddings and dynamic features. Then, we apply a communication layer using a transformer block to capture intra-vehicle \textit{dynamic} relationships:
\begin{equation}
\label{eq:transformer-layer}
\begin{aligned}
    H^\prime &= \text{Norm}(\text{MHA}(Q_t, Q_t, Q_t) + Q_t) \\
    Q^\prime_t &= \text{Norm}(\text{FFN}(H^\prime) + H^\prime)
\end{aligned}
\end{equation}
where $\text{Norm}$ denotes a normalization layer, and FFN represents the multi-layer perceptron. The self-attention in the MHA layer enables message passing between vehicles based on their current embeddings.

After this communication step, we perform cross-attention between the updated vehicle queries and the profile-specific node embeddings. Thus:
\begin{align}
    \label{eq:attn_dec}
    \text{MHA}(Q^\prime_t W^Q_i, \, \bm{h}W^K_i, \, \bm{h}W^V_i)
\end{align}
where $\bm{h} = [\bm{h}^1, \ldots, \bm{h}^m]$ are the profile-specific node embeddings from the encoder. This formulation allows the attention mechanism to consider the profile-specific encoded information for each vehicle.

The outputs of these attention heads are then combined to form $U \in \mathbb{R}^{m \times d_h}$, which is used in our Multiple Pointer Mechanism:
\begin{align}
    \label{eq:mult-pointer}
    Z = C \cdot \text{tanh} \left ( \frac{UL^\top}{\sqrt{d_h}}  \right )
\end{align}
Here, $L$ is a projection of the node embeddings $\bm{h}$, and $C$ is a scale parameter to control the entropy of $Z$  ($C=10$ in our work according to \citet{bello2016neural}). The resulting $Z \in \mathbb{R}^{m \times N}$ contains logits for each vehicle-node pair. Finally, the logit vector $Z$ of \cref{eq:mult-pointer} is masked based on the solution construction feasibility. The probability of selecting action $i$ for all vehicles is given by:
\begin{equation}
    P_i = \frac{e^{Z_i}}{\sum_i e^{Z_i}}
\end{equation}
allowing us to sample a vector $A$ containing $m$ actions for each vehicle in parallel. If a conflict arises between actions, i.e., two or more vehicles select an action of index $i$, we prioritize the vehicle with the largest action probability $P_i$ and set the action of all other conflicting vehicles equivalent to their current position as in \citet{berto2024parco}.

\subsection{Training}
\label{subsec:training}

\our{} is a centralized multi-agent parallel decision-making model, with a shared policy $\pi_\theta$ for all agents and a global reward $R$. As such, \our{} can be trained using any of the training algorithms from single-agent NCO literature. We train \our{} using the REINFORCE gradient estimator \citep{williams1992simple} with a shared baseline \citep{kwon2020pomo,kim2022sym}:

\begin{equation}
    \label{eq:reinforce}
    \nabla_\theta \mathcal{L} \approx \frac{1}{B \cdot L} \sum_{i=1}^{B} \sum_{j=1}^{L} G_{ij} \nabla_\theta \log p_\theta(A_{ij} | \bm{x}_i).
\end{equation}
Here, $B$ is the mini-batch size and $G_{ij}$ is the advantage $R(A_{ij}, \bm{x}_i) - b^{\text{shared}}(\bm{x}_i)$ of a solution $A_{ij}$ w.r.t. to the shared baseline $b^{\text{shared}}(\bm{x}_i)$ of problem instance $\bm{x}_i$, in our case obtained through symmetric transformations \citep{kim2022sym}.

For PVRP-P, different preference distributions can lead to rewards of varying scales in the reward function. To mitigate potential biases during learning, we propose applying reward balancing across these distributions, which has been successfully applied in multi-task routing problems \citep{berto2024routefinder}. We employ reward balancing techniques to calculate the normalized reward $R_{\mathrm{norm}, t}^{(k)}$ for all preference distributions $k \in [1, \cdots, K]$ at each training step $t \geq 1$. This normalization is achieved by dividing by the exponentially smoothed mean. We calculate the average reward $\hat{R}_t^{(k)}$ up to training step $t$, starting from the average reward $\bar{R}_t^{(k)}$ at step $t$. For the exponential moving average, we set $\hat{R}_1^{(k)} = \bar{R}_1^{(k)}$ and compute subsequent values for $t > 1$ based on \cite{Hunter1986}, using a smoothing factor $\beta$:
\begin{equation}
\label{eq:exponential_smoothing}
    \hat{R}_{t}^{(k)} = (1 - \beta) \cdot \hat{R}_{t-1}^{(k)} + \beta \cdot \bar{R}_{t}^{(k)}, \quad 0 < \beta < 1, \quad t > 1.
\end{equation}
We then calculate the normalized reward as $R_{\text{norm}, t}^{(k)} = R_t^{(k)} / |\hat{R}_t^{(k)}|$ to ensure fairness in reward distribution.

\section{Experiments}

\subsection{Experimental Setup}

\paragraph{Classical Baselines} We employ Google’s OR-Tools \citep{perron2023ortools}, which is both an exact and heuristic solver that utilizes constraint programming. OR-Tools is highly regarded within the operations research community for its flexibility in handling a broad spectrum of VRP variants and other problems. We also utilize PyVRP \citep{wouda2024pyvrp}, an open-source heuristic VRP solver that represents the latest advancements and is based on HGS-CVRP \citep{vidal2022hybrid}. PyVRP is capable of addressing vehicle profile constraints by accordingly modifying vehicle-wise cost matrices as we detail in \cref{sec:preliminaries}. Both baseline approaches are applied to solve each problem instance using a single CPU core.

\paragraph{Models} We evaluate the following neural methods:

\begin{itemize}
    \item \texttt{ET} \citep{son2024equity} is an advanced solution for multi-agent TSP and PDP, focusing on generating sequential actions and distributing workloads equitably.
    \item \texttt{DPN} \citep{zheng2024dpn} improves ET by using a unique encoder that improves route partitioning and navigation, significantly enhancing efficiency over previous methods.
    \item \texttt{2D-Ptr} \citep{liu20242d} dynamically adapts to varying scenarios in HCVRP using a dual-encoder system to optimize vehicle and client routes efficiently.
    \item \texttt{PARCO} \citep{berto2024parco} is a recent method that accelerates multi-agent combinatorial optimization using a novel decoding mechanism and communication layers, achieving fast and competitive results.
    \item \texttt{\our{}(-EC)} is our proposed model without the Encoder Communication (EC), i.e., without the bipartite graph message passing of \cref{subsec:encoder-camp}.
    \item \texttt{\our{}} is our full model.
\end{itemize}

For fairness of comparison, we try to match the number of training steps to be the same and adjust the batch size accordingly. Specifically, we train models for 100 epochs as in \citet{kool2018attention} using the Adam optimizer \citep{kingma2014adam} with an initial learning rate (LR) of $10^{-4}$ with a decay factor of 0.1 after the 80th and 95th epochs with the SymNCO training scheme \cite{kim2022sym}. Every epoch, we sample $10^5$ samples. The policy gradients shared baseline is made of $8$ augmented symmetrical solutions out of a $128$-batch size. We train by sampling locations and random preference scores from uniform distributions. We set the embedding dimension to $128$, the number of attention heads to $8$, and the feedforward hidden dimension to $512$ across all models and employ $3$ encoder layers. 

\begin{table*}[h!]
\centering
\caption{Benchmarks and results for PVRP-P (top) and PVRP-ZC (bottom) at varying sizes and agent numbers.  Highlighting cost ($\downarrow$) and average gaps ($\downarrow$) to the sota HGS-PyVRP. Single instance solution time in $(\cdot)$.
}
\resizebox{\textwidth}{!}{%
    \begin{tabular}[\linewidth]{l| c c c | c c c | c c c | c}
    \toprule
    \multicolumn{11}{c}{\textbf{PVRP-P (Preferences)}} \\
    \midrule
    $N$ & \multicolumn{3}{c|}{60} & \multicolumn{3}{c|}{80} & \multicolumn{3}{c|}{100} & Gap($\%$)  \\
    \midrule

    $m$ & 3 & 5 & 7 & 3 & 5 & 7 & 3 & 5 & 7 & avg.\\
        
    \midrule
    
    OR-Tools & 7.34 \small{(10m)} & 7.64 \small{(10m)} & 7.87 \small{(10m)} & 8.42 \small{(12m)} & 9.02 \small{(12m)} & 9.19 \small{(12m)} & 10.15 \small{(15m)} & 10.32 \small{(15m)} & 10.62 \small{(15m)} & 10.68 \\ 
    HGS-PyVRP     & 6.44 \small{(10m)} & 6.83 \small{(10m)} & 7.10 \small{(10m)} & 7.66 \small{(12m)} & 8.17 \small{(12m)} & 8.47 \small{(12m)} & 8.93 \small{(15m)} & 9.50 \small{(15m)} & 9.80 \small{(15m)} & 0.00 \\ 
    
    \midrule
    
    ET (\textit{g.}) & 7.62\small{(0.17s)} &8.12\small{(0.17s)} &8.41\small{(0.18s)} &9.04\small{(0.23s)} &9.63\small{(0.24s)} &9.96\small{(0.23s)} &10.50\small{(0.28s)} &11.18\small{(0.30s)} &11.63\small{(0.30s)} & 18.13\\
    DPN (\textit{g.}) & 7.52\small{(0.17s)} &8.00\small{(0.18s)} &8.27\small{(0.18s)} &8.93\small{(0.22s)} &9.63\small{(0.24s)} &10.05\small{(0.24s)} &10.50\small{(0.29s)} &11.08\small{(0.29s)} &11.42\small{(0.30s)} & 17.15\\
    2D-Ptr (\textit{g.}) & 7.34\small{(0.15s)} &7.75\small{(0.15s)} &8.03\small{(0.16s)} &8.70\small{(0.20s)} &9.30\small{(0.19s)} &9.59\small{(0.20s)} &10.10\small{(0.25s)} &10.82\small{(0.25s)} &11.18\small{(0.25s)} & 13.86\\
    PARCO (\textit{g.}) & 7.31\small{(0.14s)} &7.73\small{(0.15s)} &8.08\small{(0.15s)} &8.69\small{(0.21s)} &9.19\small{(0.22s)} &9.57\small{(0.22s)} &10.14\small{(0.25s)} &10.78\small{(0.24s)} &11.10\small{(0.25s)} & 13.21\\
    \our{}(-EC) (\textit{g.}) & 7.30\small{(0.16s)} & 7.67\small{(0.15s)} & 7.90\small{(0.16s)} & 8.68\small{(0.22s)} & 9.10\small{(0.21s)} & 9.49\small{(0.21s)} & 9.98\small{(0.25s)} & 10.72\small{(0.25s)} & 11.00\small{(0.26s)} & 12.53\\
    \our{} (\textit{g.}) & \textbf{7.16}\small{(0.17s)} &\textbf{7.66}\small{(0.18s)} & \textbf{7.88}\small{(0.18s)} & \textbf{8.49}\small{(0.25s)} & \textbf{9.07}\small{(0.25s)} & \textbf{9.44}\small{(0.25s)} & \textbf{9.87}\small{(0.33s)} & \textbf{10.60}\small{(0.33s)} & \textbf{10.90}\small{(0.33s)} & \textbf{11.23}\\
    
    \midrule
    
    ET (\textit{s.}) & 7.16\small{(0.25s)} &7.62\small{(0.24s)} &7.87\small{(0.25s)} &8.51\small{(0.35s)} &9.05\small{(0.37s)} &9.41\small{(0.36s)} &9.93\small{(0.45s)} &10.55\small{(0.44s)} &10.88\small{(0.46s)} & 11.23\\
    DPN (\textit{s.}) & 7.13\small{(0.27s)} &7.55\small{(0.26s)} &7.87\small{(0.26s)} &8.46\small{(0.34s)} &9.04\small{(0.37s)} &9.36\small{(0.36s)} &9.85\small{(0.43s)} &10.52\small{(0.42s)} &10.88\small{(0.48s)} & 10.69\\
    2D-Ptr (\textit{s.}) & 6.89\small{(0.16s)} &7.31\small{(0.17s)} &7.57\small{(0.17s)} &8.23\small{(0.20s)} &8.72\small{(0.21s)} &9.09\small{(0.22s)} &9.57\small{(0.26s)} &10.14\small{(0.26s)} &10.45\small{(0.27s)} & 6.81\\
    PARCO (\textit{s.}) & 6.87\small{(0.21s)} &7.25\small{(0.23s)} &7.55\small{(0.22s)} &8.13\small{(0.33s)} &8.66\small{(0.35s)} &9.02\small{(0.34s)} &9.44\small{(0.42s)} &10.12\small{(0.41s)} &10.49\small{(0.41s)} & 6.44\\
    \our{}(-EC) (\textit{s.}) & 6.82\small{(0.22s)} & 7.20\small{(0.23s)} & 7.54\small{(0.23s)} & 8.12\small{(0.33s)} & 8.63\small{(0.32s)} & 8.95\small{(0.34s)} & 9.42\small{(0.40s)} & 10.05\small{(0.42s)} & 10.32\small{(0.41s)} & 5.72\\
    \our{} (\textit{s.}) &\textbf{ 6.75}\small{(0.33s)} &\textbf{7.18}\small{(0.35s)} &\textbf{7.45}\small{(0.33s)} &\textbf{8.05}\small{(0.43s)} &\textbf{8.57}\small{(0.42s)} &\textbf{8.90}\small{(0.42s)} &\textbf{9.38}\small{(0.51s)} &\textbf{9.96}\small{(0.54s)} &\textbf{10.29}\small{(0.54s)} & \textbf{5.02}\\
    
    \bottomrule
    \end{tabular}
    }
    \\
    \resizebox{\textwidth}{!}{%
    \begin{tabular}[\linewidth]{l| c c c | c c c | c c c | c}
    \multicolumn{11}{c}{\textbf{PVRP-ZC (Zone Constraints)}} \\
    \midrule
    $N$ & \multicolumn{3}{c|}{60} & \multicolumn{3}{c|}{80} & \multicolumn{3}{c|}{100} & Gap($\%$)  \\
    \midrule

    $m$ & 3 & 5 & 7 & 3 & 5 & 7 & 3 & 5 & 7 & avg.\\
        
    \midrule
    
    OR-Tools & 13.17\small{(10m)} & 13.15\small{(10m)} & 13.18\small{(10m)} & 16.67\small{(12m)} & 16.69\small{(12m)} & 16.70\small{(12m)} & 20.18\small{(15m)} & 20.22\small{(15m)} & 20.20\small{(15m)} & 8.35\\
    HGS-PyVRP & 12.13\small{(10m)} & 12.16\small{(10m)} & 12.16\small{(10m)} & 15.36\small{(12m)} & 15.40\small{(12m)} & 15.42\small{(12m)} & 18.59\small{(15m)} & 18.66\small{(15m)} & 18.70\small{(15m)} & 0.00\\
    
    \midrule
    
    ET (\textit{g.}) & 13.87\small{(0.17s)} & 14.04\small{(0.17s)} & 13.91\small{(0.17s)} & 17.55\small{(0.23s)} & 17.54\small{(0.23s)} & 17.48\small{(0.24s)} & 21.28\small{(0.28s)} & 21.20\small{(0.29s)} & 21.36\small{(0.29s)} & 14.15\\
    DPN (\textit{g.}) & 13.98\small{(0.17s)} & 13.94\small{(0.17s)} & 13.81\small{(0.17s)} & 17.50\small{(0.22s)} & 17.62\small{(0.22s)} & 17.57\small{(0.23s)} & 21.24\small{(0.28s)} & 21.31\small{(0.28s)} & 21.37\small{(0.28s)} & 14.14\\
    2D-Ptr (\textit{g.}) & 13.42\small{(0.15s)} & 13.49\small{(0.15s)} & 13.37\small{(0.15s)} & 16.98\small{(0.20s)} & 16.95\small{(0.20s)} & 17.04\small{(0.20s)} & 20.53\small{(0.25s)} & 20.58\small{(0.24s)} & 20.57\small{(0.25s)} & 10.06\\
    PARCO (\textit{g.}) & 13.39\small{(0.14s)} & 13.42\small{(0.13s)} & 13.32\small{(0.13s)} & 16.87\small{(0.21s)} & 16.95\small{(0.20s)} & 16.82\small{(0.20s)} & 20.33\small{(0.25s)} & 20.52\small{(0.25s)} & 20.33\small{(0.25s)} & 9.55\\
    \our{}(-EC) (\textit{g.}) & 13.05\small{(0.17s)} & 13.19\small{(0.16s)} & \textbf{13.05}\small{(0.16s)} & \textbf{16.58}\small{(0.22s)} & 16.73\small{(0.23s)} & 16.65\small{(0.22s)} & 20.10\small{(0.25s)} & \textbf{20.07}\small{(0.25s)} & 20.07\small{(0.26s)} & 7.69\\
    \our{} (\textit{g.}) & \textbf{12.93}\small{(0.19s)} & \textbf{13.07}\small{(0.20s)} & 13.14\small{(0.18s)} & 16.59\small{(0.25s)} & \textbf{16.38}\small{(0.25s)} & \textbf{16.52}\small{(0.24s)} & \textbf{20.00}\small{(0.33s)} & 20.18\small{(0.32s)} & \textbf{20.07}\small{(0.33s)} & \textbf{7.62}\\
    
    \midrule
    
    ET (\textit{s.}) & 13.18\small{(0.25s)} & 13.17\small{(0.25s)} & 13.18\small{(0.25s)} & 16.63\small{(0.34s)} & 16.73\small{(0.34s)} & 16.58\small{(0.34s)} & 20.19\small{(0.46s)} & 20.47\small{(0.45s)} & 20.14\small{(0.45s)} & 8.78\\
    DPN (\textit{s.}) & 13.13\small{(0.27s)} & 13.30\small{(0.27s)} & 13.24\small{(0.26s)} & 16.62\small{(0.33s)} & 16.67\small{(0.33s)} & 16.78\small{(0.34s)} & 20.23\small{(0.44s)} & 20.18\small{(0.44s)} & 20.34\small{(0.43s)} & 8.69\\
    2D-Ptr (\textit{s.}) & 12.86\small{(0.15s)} & 12.93\small{(0.15s)} & 12.99\small{(0.15s)} & 16.33\small{(0.21s)} & 16.31\small{(0.21s)} & 16.17\small{(0.22s)} & 19.76\small{(0.25s)} & 19.72\small{(0.25s)} & 19.90\small{(0.25s)} & 5.92\\
    PARCO (\textit{s.}) & 12.85\small{(0.21s)} & 12.81\small{(0.22s)} & 12.86\small{(0.22s)} & 16.28\small{(0.33s)} & 16.46\small{(0.33s)} & 16.31\small{(0.33s)} & 19.64\small{(0.42s)} & 19.72\small{(0.42s)} & 19.79\small{(0.41s)} & 5.86\\
    \our{}(-EC) (\textit{s.}) & \textbf{12.49}\small{(0.22s)} & 12.58\small{(0.23s)} & 12.56\small{(0.23s)} & 15.87\small{(0.33s)} & 15.92\small{(0.32s)} & \textbf{15.86}\small{(0.33s)} & 19.20\small{(0.41s)} & \textbf{19.19}\small{(0.41s)} & 19.18\small{(0.40s)} & 3.34\\
    \our{} (\textit{s.}) & 12.54\small{(0.33s)} & \textbf{12.50}\small{(0.33s)} & \textbf{12.52}\small{(0.34s)} & \textbf{15.77}\small{(0.42s)} & \textbf{15.84}\small{(0.43s)} & 15.88\small{(0.42s)} & \textbf{19.06}\small{(0.50s)} & 19.26\small{(0.49s)} & \textbf{19.18}\small{(0.49s)} & \textbf{3.31}\\    
    
    \bottomrule
    \end{tabular}
    }
\label{table:pvrp-zc}
\end{table*}

\paragraph{Environment Settings} We evaluate the trained models under four types of environment settings:

\begin{itemize}
    \item \texttt{Random}: each vehicle has a random preference score between $0$ and $1$ for each client.
    \item \texttt{Angle}: based on the depot's location, the region is divided into sectors by angle. Each vehicle is randomly assigned to one sector, and all clients within that sector are given a preference score of $1$ for this vehicle, while clients outside the sector have a score of $0$.
    \item \texttt{Cluster}: a number of vehicle cluster centers are randomly placed in the region. Each vehicle is assigned to one cluster. The preference score for each vehicle for each client is the Euclidean distance from the client to the respective cluster center.
    \item \texttt{Zone}: cluster centers are randomly placed in the region from the number of vehicles to three times that number. Each client belongs to the zone of the nearest cluster center. Each zone is randomly designated as available or unavailable to each vehicle.
\end{itemize}

\paragraph{Evaluation Settings} We randomly generate $1280$ instances for each type of environment setting, with varying numbers of vehicles and clients. All baselines are evaluated across settings of $\alpha$ ranging from $0.0$ to $0.2$ for PVRP-P. We calculate the average cost among different preference distributions as the final cost for PVRP-P.

Evaluation runs are conducted on an AMD Ryzen Threadripper 3960X 24-core CPU with a single RTX 3090 GPU. We value open reproducibility and provide source code on Github \footnote{\url{https://github.com/ai4co/camp}}.

\subsection{Experiment Results}

\begin{figure*}[h!]
    \centering
    \includegraphics[width=\linewidth]{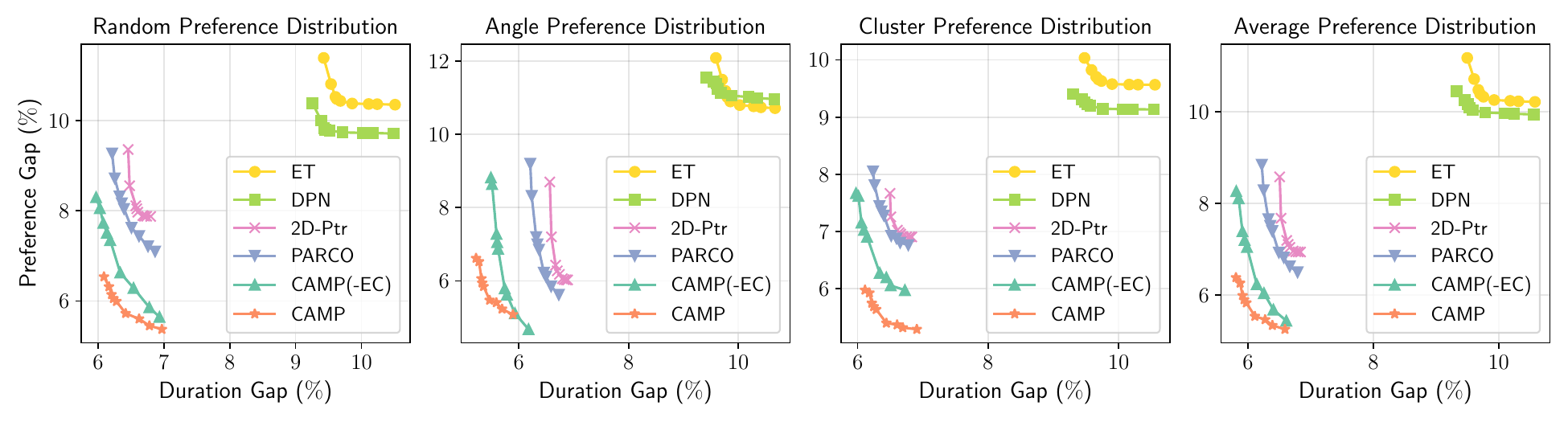}
    \caption{Pareto plot of VRP-P gaps at varying values of $\alpha$ for different preference matrix distribution. The bottom left is better.}
    \label{fig:pareto}
\end{figure*}

\begin{figure}[h!]
    \centering
    \includegraphics[width=0.95\linewidth]{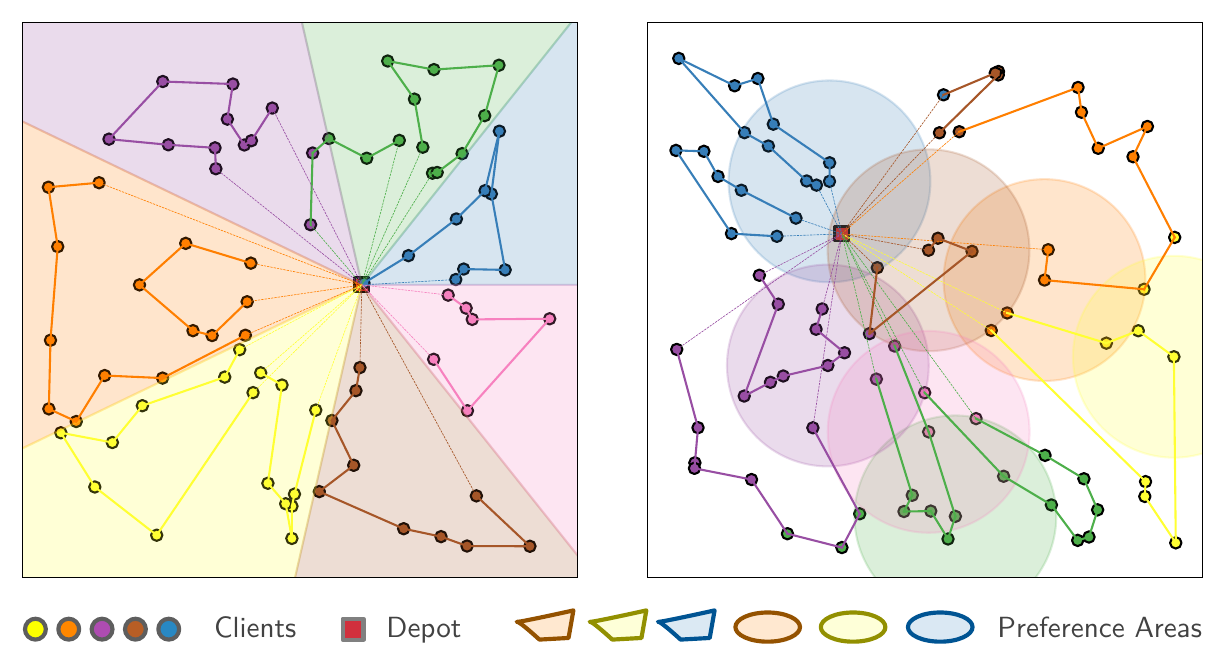}
    \caption{Visualization of \our{} solutions of PVRP-P with angle and cluster preference distribution. \our{} successfully routes according to the preference distribution.}
    \label{fig:visualization}
\end{figure}



\cref{table:pvrp-zc} compares \our{} against previously discussed baselines for PVRP-P and PVRP-ZC, with inference times shown in parentheses $(\cdot)$; with (\textit{g.}) referring to greedy performance while (\textit{s.}) refers to sampling 1280 solutions. We observe that \our{} consistently outperforms all other neural solver baselines in experiments across all client and vehicle sizes for both PVRP-P and PVRP-ZC. \cref{fig:pareto} displays the Pareto curve for all models at various $\alpha$ values under different preference distributions. The higher the $\alpha$, the greater the weight given to preferences relative to costs. It is evident that \our{} offers a significant advantage in optimizing both duration and preference compared to all other baselines across all preference distributions. As the weight of the preference increases, \our{} improves more effectively in optimizing preference compared to other models while maintaining robust performance in duration. This indicates our model's superior capability in capturing vehicle preference information.

\cref{fig:visualization} qualitatively visualizes the solutions constructed by \our{} for instances of PVRP-P with angle and cluster preference distributions. Each color represents a vehicle's preferred area and its corresponding route. It is noteworthy that \our{} effectively constructs solutions that respect the preferences of vehicles while maintaining optimal duration. This results in a superior solution that optimizes both duration and preference.

\subsection{Ablation Studies}

\cref{tab:camp-results-ablation}
Illustrates the main ablation studies for the \our{} components on PVRP-P with instance size $N=100$. We compare the performance of the full \our{} model against its variants with specific components ablated one at a time.

\begin{table}[h!]
\caption{Ablation study for CAMP model showing gaps $(\downarrow)$ removing different components.}
\label{tab:camp-results-ablation}
\centering
\begin{tabular}{@{}lcc@{}}
\toprule
\textbf{Model} & \textbf{Dur. Gap} & \textbf{Pref. Gap} \\
\midrule
CAMP (full) & \textbf{6.52\%} & \textbf{4.98\%} \\
- Encoder Communication & 6.55\% & 5.68\% \\
\quad - Balanced Reward Training & 6.63\% & 5.88\% \\
\qquad - Vehicle-specific Profile Embedding & 6.85\% & 6.35\% \\
\bottomrule
\end{tabular}
\end{table}

\paragraph{Encoder Communication} This ablation (\our{}-EC in \cref{table:pvrp-zc}) underpins the effectiveness of encoder communication due to the bipartite graph, which significantly enhances the representation capability of \our{} in capturing the preference relationships, leading to improved optimization of the preference gap.

\paragraph{Balanced Reward Training} Comparing \our{} without Encoder Communication and Balanced Reward, we observe that the latter successfully equilibrates the rewards across different preference settings, making it easier for the model to adapt and thereby enhancing performance in managing preferences.

\paragraph{Vehicle-specific Profile Embedding} This setting removes all our contributed components in \our{} and corresponds to PARCO \citep{berto2024parco}. This overall performs worst, although still better than other baselines such as the sequential autoregressive ones.


\section{Conclusion}
\label{sec:conclusion}

In this work, we introduced a formulation for the Profiled Vehicle Routing Problem (PVRP), an extension of the Heterogeneous Capacitated Vehicle Routing Problem (HCVRP) that incorporates client-specific preferences and operational constraints. To tackle this complex problem, we proposed the Collaborative Attention Model with Profiles (\our{}), a novel multi-agent reinforcement learning (MARL) approach that leverages an attention-based encoder-decoder framework with agent communication to enable collaborative decision-making among heterogeneous vehicles with different profiles for each client. Our extensive evaluations on both synthetic, across two types of PVRP variants—PVRP with Preferences (PVRP-P) and PVRP with Zone Constraints (PVRP-ZC)—show that \our{} consistently delivers competitive performance in terms of solution quality and computational efficiency, outperforming traditional heuristics and other neural-based methods. These results position \our{} as a powerful tool for solving complex routing problems in dynamic, real-time settings.

\paragraph{Limitations and Future Work.} \our{} represents an early attempt at solving the PVRP. While it shows promising results in solution time and performance -- including outperforming OR-Tools -- it can be improved in several ways to beat the final performance of SOTA heuristic HGS. Promising future works include integrating end-to-end construction and improvement methods \citep{kong2024efficient}, learning to guide (local) search algorithms \citep{hottung2020neural,li2021learning,yan2024neural}, multi-objective learning at different preference values $\alpha$ \citep{lin2022pareto_pmoco,chen2024neural_nhde}, extending recent foundation models for VRPs \citep{berto2024routefinder,li2024cada} with \our{}'s \textit{agentic} representations including agent communication and heterogenous learned representations, 
and obtaining better heuristics for resolving decoding conflicts which could be achieved by automated LLM algorithmic discovery \citep{liu2024evolution,ye2024reevo,pham2025hsevo}.






\bibliographystyle{ACM-Reference-Format} 
\bibliography{bibliography}

\end{document}